INVITED REVIEW

# Thermal transport in semiconductor nanostructures, graphene and related two-dimensional materials

Alexandr I. Cocemasov, Calina I. Isacova and Denis L. Nika[1]

E. Pokatilov Laboratory of Physics and Engineering of Nanomaterials, Department of Physics and Engineering, Moldova State University, Chisinau MD-2009, Republic of Moldova

**Abstract**

We review experimental and theoretical results on thermal transport in semiconductor nanostructures (multilayer thin films, core/shell and segmented nanowires), single- and few-layer graphene, hexagonal boron nitride, molybdenum disulfide and black phosphorus. Different possibilities of phonon engineering for optimization of electrical and heat conductions are discussed. The role of the phonon energy spectra modification on the thermal conductivity in semiconductor nanostructures is revealed. The dependence of thermal conductivity in graphene and related two-dimensional (2D) materials on temperature, flake size, defect concentration, edge roughness and strain is analyzed.

**Keywords:** phonons, thermal conductivity, nanowire, graphene, 2D materials

**PACS:**   63.22.Rc, 63.22-m, 65.80.Ck, 65.80.-g

---

[1] Corresponding author (DLN): dlnika@yahoo.com

## 1. *Thermal transport in semiconductor nanostructures*

Rapid miniaturization of electronic devices to nanoscale range requires new approaches for efficient management of their heat and electrical conductions. One of these approaches, referred to as *phonon engineering* [1], is related to optimization of thermal and electronic properties of nanodimensional structures due to modification of their phonon properties [1-3]. At the end of the previous century several research groups have demonstrated that many phonon confined branches appear in energy spectra of homogeneous semiconductor thin films and nanowires [4-9], leading to change in phonon density of states and reduction of average phonon group velocity in comparison with corresponding bulk materials [7-9]. The latter together with enhanced phonon boundary scattering results in decreasing of lattice thermal conductivity (TC). Balandin and Wang [7] have theoretically predicted that lattice thermal conductivity of 10-nm-wide silicon film is by an order of magnitude smaller than that in bulk silicon at room temperature (RT). Five-times drop of lattice thermal conductivity was also theoretically predicted for Si nanowire with a diameter of 20 nm [10]. Subsequent independent theoretical studies [11-17] and experimental measurements of thermal conductivity in several nm-thick free-standing Si films and nanowires [18-21] confirmed the initial predictions: strong reduction of lattice thermal conductivity as compared with bulk material was revealed.

More precise tuning of phonon properties and heat conduction at nanoscale can be realized in multilayer films (MFs) and core/shell nanowires (NWs) [22-33]. The evolution of phonon energies in homogeneous silicon films and silicon films covered by diamond claddings is illustrated in Figure 1, where we show the dispersion relations for the dilatation (SA) phonon modes in the freestanding Si film (a); Diamond/Si/Diamond heterostructures with the different thickness of the diamond (D) barrier layer (b-c); and Si film with the clamped external surfaces (d), which correspond to a film embedded in the "absolutely" rigid material. The thickness of the Si layer in all cases is 2 nm to insure the strong phonon confinement effect at RT. The phonon energies for each phonon branch $s$ in the freestanding film are lower than those in the clamped film. Phonon modes of D/Si/D heterostructure occupy an intermediate position between two limiting cases depending on the diamond layer thickness.

<Figure 1>

Modification of phonon energy spectra and phonon density of states in MFs and core/shell NWs leads to significant change in average phonon group velocity $\bar{v}^{(\alpha)}$. The dependence of $\bar{v}^{(\alpha)}$ on the phonon frequency for dilatation phonon modes in bare GaN nanowire, GaN/AlN and GaN/Plastic core/shell nanowires is presented in Figure 2. The average group velocity curves $\bar{v}(\omega)$ are strongly oscillating functions due to the presence of many quantized phonon branches $s$ and the fast variation of the derivatives $1/v_s^{(\alpha)}(\omega)) = dq_s^{(\alpha)}(\omega)/d\omega$ with changing $\omega$. Acoustically soft and slow plastic shell decreases the phonon group velocity by a factor of 3 in comparison with bare GaN NW, while acoustically hard and fast AlN shell leads to an opposite effect of increasing $\bar{v}^{(\alpha)}$ by a factor of 1.3 [24].

<Figure 2>

The effect of shell layers on thermal conductivity of Si film is illustrated in Figure 3, where temperature dependence of thermal conductivity is shown for bare Si NW and core/shell Si/Diamond, Si/Plastic and Si/SiO$_2$ NWs. Diamond shell with higher sound velocity than that in Si increases thermal conductivity in core/shell NWs in comparison with Si NW at temperatures $T$>150 K. Plastic and SiO$_2$ shells (with low sound velocity) suppress heat propagation in core/shell nanowires. Change in the thermal conductivity is primarily induced by the mixing of phonons from the core and the shell, leading to appearance of hybrid phonon modes which propagate through the whole core/shell nanowire.

<Figure 3>

Segmented nanowires consisted from segments of different materials and shapes were recently proposed as efficient phonon filters, removing many phonon modes from the heat flux due to their trapping in nanowire segments [17,27,30,33]. In Figure 4 is presented the temperature dependence of phonon thermal conductivity for Si NW, segmented cross-section modulated Si NW (SMNW) and segmented cross-section modulated Si/Ge core/shell nanowires. Increasing the thickness of Ge shell from 1 ML to 7 ML leads to a decrease of thermal conductivity of Si/Ge SMNW by a factor of 2.9 – 4.8 in comparison with that in Si SMNW without Ge shell, and by a factor of 13 – 38 in comparison with that in bare Si NW. The reduction in $\kappa$ of Si/Ge segmented cross-section modulated NWs is substantially stronger than that reported for core/shell nanowires without cross-sectional modulation [16, 28-29, 31]. In core/shell nanowires without modulation the drop of thermal conductivity occurs mainly due to hybridization of phonon modes, resulting in changes of PDOS and phonon relaxation. In SMNWs, reduction of lattice thermal conductivity is reinforced due to interplay between hybridization of phonon modes and their localization in NW segments, which completely removes such modes from the heat flux [33].

<Figure 4>

Spatial confinement of phonons in nanostructures opens up new possibilities for engineering of electron – phonon interaction and electron mobility. Formation of electron and phonon minibands in nanostructures with energy gaps between confined branches may result in phonon bottleneck effect [34-36], non-monotonic (oscillated) dependence of electron mobility on nanostructure thickness [37-40], suppression of electron-phonon scattering at certain electron energies and enhancement of electron mobility [25, 41-43]. Nika et al. [25] and Fonoberov et al. [43] have shown theoretically that RT electron mobility in Si thin films or Si nanowires with diamond barrier layers is by a factor of 2 – 4 higher than that in bare Si films or NWs. The enhancement effect was explained by strong modification of phonon energy spectra inside the Si channel layer, leading to reduction of electron-phonon scattering [25, 43].

Although the appearance of confined phonon energy branches in semiconductor nanostructures has been predicted theoretically by many research groups and has been indirectly confirmed by thermal conductivity measurements, direct observation of confined acoustic phonon modes was made very recently, using Brillouin-Mandelstam light scattering spectroscopy [44]. Figure 5 shows the acoustic phonon energy branches in GaAs nanowires with diameter of 122 nm. Solid lines are calculated in the framework of elastic continuum approach, while open dots represent experimental phonon energies, measured using Brillouin-Mandelstam spectrometer. Confined phonon energy branches separated by energy gaps are clearly seen in Figure 5.

<Figure 5>

## 2. *Thermal transport in single-layer graphene*

Monoatomic sheet of sp$^2$-hybridized carbon atoms – graphene demonstrates unique electrical [45-46], optical [47-48] and thermal [49-51] properties owing to quasi two-dimensional electron and heat propagation. Extremely high values of thermal conductivity [49-52] and electron mobility [53-54] make graphene a promising material for heat-removal applications and electronics. First experimental measurements of thermal conductivity in suspended single-layer graphene (SLG) flakes were performed using Raman optothermal method, developed by Balandin and co-workers [49, 52]. In this method, laser light was used for the heating-up graphene flakes suspended over trench, while temperature rise was extracted from the shift of the Raman G-band. The ultra-high values of RT thermal conductivity κ ~ 3000 – 5000 W/m-K were measured in rectangular graphene flakes with dimensions 5 μm ×10 μm [49, 52]. These values are by an order of magnitude higher than thermal conductivity of cuprum or aurum and by a factor of 2 – 3 higher than thermal conductivity of bulk diamond or highly-oriented pyrolytic graphite. A significant experimental and theoretical efforts have been made later to investigate peculiarities of phonon modes and thermal transport in graphene and graphene nanoribbons [55-80]. The thermal conductivity of suspended graphene in the range 1600 – 3800 W/m-K at temperatures close to the RT was measured in Refs. [55-59], using different experimental techniques: Raman optothermal [55-58] or electrical self-heating [59] methods. Different theoretical models were employed for the investigation of two-dimensional phonon transport in graphene: Boltzmann-transport equation (BTE) approach [61-68,73-75], equilibrium and non-equilibrium molecular dynamics [71,72,76-78]. Strong dependence of thermal conductivity on temperature, flake size and shape, strain, edge quality and crystal lattice defects was predicted in numerous theoretical studies [61-79]. The phonon mean free path (MFP) in high-quality suspended graphene was estimated to be as long as 800 nm at RT [52]. Therefore, thermal conductivity is very sensitive to both intrinsic and extrinsic phonon scatterings. The latter is crucial for the engineering of phonon and thermal properties of graphene. The dependence of lattice thermal conductivity in singe-layer graphene on temperature and flake width *d* is illustrated in Figure 6 (a). The theoretical curves were calculated in the framework of BTE approach with taking into account three-phonon Umklapp and edge roughness scatterings. The thermal conductivity decreases with temperature rise due to enhancement of three-phonon

Umklapp scattering. However, even at $T = 400$ K, the influence of flake width on κ is very pronounced: decreasing $d$ from 9 μm to 3 μm results in twofold reduction of thermal conductivity. At the same time, the four-phonon scattering limits MFP of low-energy phonons, removes the dependence of κ on flake length $L$ for $L>100$ μm [67] and may decrease the lattice thermal conductivity [80]. Dependence of RT TC on defect concentration induced by electron beam irradiation is presented in Figure 6(b). Solid curves correspond to theoretical results calculated using the BTE with different values of specularity parameter $p$. The experimental data are shown by squares, circles and triangles. For the small defect densities $N_D < 1.2 \times 10^{11}$ cm$^{-2}$, the thermal conductivity rapidly decreases with $N_D$. At higher defect density the thermal conductivity revealed weakening of the κ($N_D$) dependence, which was explained theoretically by the interplay between the three main phonon scattering mechanisms – Umklapp scattering due to lattice anharmonicity, mass-difference scattering, and rough edge scattering [78].

<Figure 6>

Thermal conductivity of graphene supported on a substrate [60] or graphene encased within cladding layers [81] is several times smaller than that in suspended graphene due to phonon scattering on interfaces between graphene and substrate/cladding layers and modification of graphene phonon modes. The experimental and theoretical data on the thermal conductivity of single-layer graphene are summarized in Tables I and II. Readers interested in more detailed description of phonon modes and thermal transport in graphene are referred to different reviews [2,50,51,82-87]. Application of Raman optothermal method for the thermal conductivity measurements in graphene and related two-dimensional materials is comprehensively described in the recent review by Malekpour and Balandin [88].

### 3. *Thermal transport in few-layer graphene*

Several interesting results have been obtained in the field of thermal transport in few-layer graphene (FLG) during the last few years [89-104]. A series of independent measurements of thermal conductivity in FLG has been performed in Ref. [89-92]. Ghosh et al. [89] reported on the measurements of thickness-dependent thermal conductivity in FLG. Rapid decrease of TC with increase of number of layers from 1 up to 4 was observed and explained theoretically by increase in phase space allowed for three-phonon Umklapp scattering [89]. Jo et al. [90] measured the TC in exfoliated and suspended bilayer graphene (BLG) samples using electro-thermal micro-bridge method and found TC in the range $(730 – 880) \pm 60$ W/m-K at RT. These values are 2 - 3 times lower than those obtained using Raman optothermal technique [58,89], while in accordance with previous electro-thermal measurements [93-95]. It was argued that this suppression is mainly caused by scattering from polymer residues on the graphene surface. Li et al. [58] employed Raman optothermal technique for investigation of thermal transport in twisted bilayer graphene (T-BLG). The authors found that in the wide range of temperatures 300 - 750 K the TC in T-BLG is smaller than both in SLG and AB-BLG (see Fig. 7(a)). The thermal conductivity of T-BLG is by a factor of ~ 2 smaller than that in SLG and by a factor of ~ 1.35 smaller than that in AB-BLG near the RT. The drop of TC was explained by emergence of many additional hybrid folded phonons in T-BLG resulting in more intensive phonon scattering [58, 96-98]. Jeong et al. [91] reported the TC of $2868 \pm 932$ W/m-K at RT for CVD large-scale free-standing FLG in which a micropipette temperature sensor with an inbuilt laser point heating source was used. The large uncertainty of 32 % in TC measurement was attributed mainly due to the non-uniform thickness of the film in the range $1 – 10$ layers with an average thickness around 7 layers. The layer numbers were determined from the intensity ratio of Raman G-band to 2D-band and full width at half maximum of 2D-band. Liu et al. [92] have measured the TC at RT of millimeter-size CVD FLG supported on organic substrate (PMMA) in the range 33 - 365 W/m-K depending on the FLG thickness, using the transient electro-thermal technique. The mm-scale of the sample eliminated the thermal contact resistance problems and phonon-edge scattering encountered in µm-scale samples. The obtained values are about one order of magnitude lower than TC of suspended graphene and are significantly lower than TC of supported graphene on $SiO_2$. The authors assumed that the abundant C atoms in the PMMA enhance the energy and momentum exchange with the supported graphene, thus reinforcing the phonon scattering and reducing the TC of CVD graphene on PMMA in comparison with $SiO_2$ substrate.

Nika et al. [97] and Cocemasov et al. [96,98] have theoretically demonstrated the possibility of phonon engineering of thermal properties of BLG and other 2D layered materials by twisting (rotating) the atomic planes. In these studies [96-98] the calculations of phonon energy spectra in T-BLG were performed using the Born-von Karman model of lattice dynamics for intralayer atomic interactions and spherically symmetric interatomic potential for interlayer interactions. Nika et al. [97] revealed an intriguing dependence of the specific heat $c_V$ in T-BLG on the rotational angle $\theta$, which is particularly pronounced at low temperatures (see Fig. 7(b)). Specifically, the twisting decreased the specific heat in T-BLG by 10 % – 15 % in comparison with AB-BLG at $T = 1$ K. Cocemasov et al. [98] within a more rigorous model, taking into account both anisotropy of phonon dispersions and non-parabolicity of ZA branch, have shown that the low-temperature specific heat scales with temperature as $c_V \sim T^{1.3}$ for AB-BLG and $c_V \sim T^{1.6}$ for T-BLG with rotation angle $\theta = 21.8°$.

<Figure 7>

Limbu et al. [99] reported on the measurement of the RT TC of CVD synthesized polycrystalline T-BLG as a function of grain size employing a noncontact optical technique based on micro-Raman spectroscopy. The measured TC values are $1305 \pm 122$ W/m-K, $971 \pm 73$ W/m-K and $657 \pm 42$ W/m-K for average grain sizes of 54 nm, 21 nm, and 8 nm, respectively. The equilibrium molecular dynamics (EMD) simulations with reactive empirical bond order potential for covalent intralayer interaction and Lennard-Jones potential for interlayer interaction have also been employed [99]. The calculations revealed that the degradation in TC due to grain boundaries is smaller in a polycrystalline BLG than in a polycrystalline SLG, which resulted from non-negligible interactions between adjacent graphene layers. Kuang et al. [100] have calculated the RT TC of FLG up to four layers and graphite under different isotropic tensile strains in the framework of density functional theory (DFT) combined with phonon Boltzmann-Peierls equation. Enhancement of the intrinsic TC up to 40 % was found both for FLG and graphite with strain amplification. It was concluded that this TC behavior was determined by competition between the decreased mode heat capacities and the increased lifetimes of flexural phonons. A similar TC behavior was observed for 2-layer hexagonal boron nitride systems. These results provide insights into engineering of the TC of FLG and related 2D layered materials by strain. Zhan et al. [101] have investigated the thermal transport properties of BLG with interlayer linkages using the large-scale nonequilibrium molecular dynamics (NEMD) simulations and an adaptive intermolecular reactive empirical bond order potential. It was found that the BLG TC could be effectively tailored through the introduction of different types of interlayer linkages, i.e. divacancy bridgings, "spiro" interstitial bridgings and Frenkel pair defects. The impacts from the interlayer linkages could be further modulated through their density and distribution. An up to 80 % reduction in the TC was predicted for the BLG with 1.39 % divacancy bridgings. It was also found that the linkages that contain vacancies (divacancy bridgings and Frenkel pair defects) lead to more severe suppression of the TC owing to a stronger phonon scattering. Shen et al. [102] have studied the effects of number of graphene layers $n$ and their size on TC of FLG embedded in an epoxy matrix. Their NEMD-based simulations have shown that the in-plane TC of FLG and the TC across the FLG/epoxy interface simultaneously increase with increasing number of graphene layers. However, higher TC of FLG is not translated into higher TC of graphene/epoxy bulk composites unless they have large lateral sizes to maintain their aspect ratios comparable to the few-layer counterparts [102]. These findings signify the strong influence of FLG lateral sizes on its TC and offer a simple approach to effective improvement of TCs of composites using FLG sheets.

Si et al. [103] employed the EMD simulations to study the effects of torsion on the TC of FLG. TC of 10-layer FLG with torsion angles of 0, 11.25, 22.5, 33.75, 45, 67.5, 90, 112.5, and 135 degrees was calculated. The simulations revealed an intriguing dependence of TC on torsion angles, namely, there is a minimum at torsion angle 22.5 degrees with the lowest TC of 4692 W/m-K. The torsion effect was analyzed as a combination of the compression effect and the dislocation effect. The spectral energy density analysis showed that the effect of dislocation on TC can be negligible, while the compression effect decreases the phonon lifetimes of flexural ZA branches and increases the ZA group velocities and the phonon specific heat. The decrease dominated when the torsion angle was small, whereas the increase dominated at large torsion angles, which are responsible for the reported variation of TC. D'Souza and Mukherjee [104] reported on the calculation of the TC of SLG and BLG from DFT and BTE approach. An excellent agreement with experimental data in the temperature range 300–700 K has been achieved in Ref. [58].

## 4. Thermal transport in related 2D layered materials

### 4.1. Hexagonal boron nitride

Hexagonal boron nitride (h-BN) holds many physical properties of practical importance such as high thermal conductivity, temperature stability and excellent impermeability [105]. However, compared to graphene thermal field, there are much less experimental data for the TC of h-BN, although several interesting reports appeared in recent years [106-108]. Zho et al. [107] using the optothermal Raman technique have measured the TC in suspended 9-layer h-BN sheets synthesized by a low pressure CVD method. They found TC in the range 227 - 280 W/m-K at RT, which is close to the data previously reported for 5-layer h-BN [106] and comparable to that of bulk h-BN (~ 390 W/m-K) [109]. More recently Wang et al. [108] provided experimental observation of the thickness-dependent TC in suspended few-layer h-BN similar to that of few-layer graphene. They have found RT TC of a suspended bilayer h-BN sample to be around 484 W/m-K which clearly exceeds that of bulk h-BN, but smaller than theoretically predicted value of 520 W/m-K in single layer h-BN [110].

### 4.2. Molybdenum disulfide

Molybdenum disulfide ($MoS_2$), one of the transition metal dichalcogenides, has a natural thickness-depended energy gap [111-112], which makes it promising for electronic and optoelectronic applications. Recently there have been several experimental [113-115] and theoretical [116-119] studies on the thermal transport of single- and few-layer $MoS_2$. Yan et al. [113] reported on a Raman characterization of exfoliated and suspended single layer $MoS_2$. The RT TC was found to be 34.5 ± 4 W/m-K, which is by a factor of 3 lower than that of bulk $MoS_2$ [120] and by ~ 2 orders of magnitude lower than that of single layer graphene [49]. A BTE-based calculation from the same work [113] agreed well with the measured value. Jo et al. [114] used the electro-thermal micro-bridge method to obtain the TC of suspended few-layer $MoS_2$ as a function of temperature. The measured RT TC values were 44 – 50 W/m-K and 48 – 52 W/m-K for 4- and 7-layer samples, respectively. In contrast to the suspended sample measurements, a relatively high TC of 62 W/m-K at RT was reported by Taube et al. [115] for supported single layer $MoS_2$ using Raman optothermal technique. The authors suggested that the discrepancy may be associated with the different sample preparation and measurement conditions. Cai et al. [116] using the density functional perturbation theory (DFPT) have investigated the vibrational and thermal properties of single layer MoS$_2$. It was found that the dominated phonon mean free path is less than 20 nm, about 30-fold smaller than that of graphene. Combined with the nonequilibrium Green's function calculations the RT TC around 23 W/m-K was predicted. Consistently, a value around 20 W/m-K of RT TC in single layer MoS$_2$ was calculated by Ding et al. [117] based on NEMD simulations. In another study by the same authors [118] the effects of lattice defects and mechanical strain on the TC of single layer MoS$_2$ have been investigated. It was predicted that a 0.5 % concentration of mono-Mo vacancies is able to reduce the TC by about 60 %. Moreover, a 12 % tensile strain is able to reduce the TC by another 60 %. A theoretical calculation by Gu et al. [119] based on phonon Boltzmann-Peierls equation approach has shown a decreasing trend in TC while going from 1- to 3-layer MoS$_2$ due to the changes in phonon dispersion relations and higher phonon scattering rates induced by the increased anharmonicity in few-layer MoS$_2$.

*4.3. Black phosphorus*

Single- and few-layer black phosphorus (phosphorene) is another promising 2D material for future electronics [121-123]. Due to its two distinct high-symmetry directions, denoted as zigzag (ZZ) and armchair (AC), phosphorene demonstrates anisotropic properties which could be useful for an efficient transport management. Recently TC measurements in single- and few-layer phosphorene were reported by three different groups using Raman optothermal, thermal bridge and time-domain thermoreflectance methods, respectively [124-126]. The measured values showed a significant anisotropy ranging from $(10 - 34) \pm 4$ W/m-K in ZZ direction and $(17 - 86) \pm 8$ W/m-K in AC direction, depending on the thickness of the films. Theoretical modeling from Ref. [124] revealed that the observed anisotropy is primarily related to the anisotropic phonon dispersion, whereas the intrinsic phonon scattering rates were found to be similar along the ZZ and AC directions. Hong et al. [127] using NEMD simulations have predicted RT TC for infinite length ZZ and AC single layer phosphorene to be $110.7 \pm 1.75$ W/m-K and $63.6 \pm 3.9$ W/m-K, respectively. The thermal transport anisotropy was partially attributed to the direction-dependent phonon group velocities. Liu and Chang [128] employing BTE approach have found that TC in single layer phosphorene is a smoothly decreasing function of the crystal chirality while going from ZZ to AC directions (see Fig. 8). An unusually high contribution of 30 % to the total TC from optical phonons was also predicted due to the group velocities and relaxation times comparable to those of acoustic phonons.

<Figure 8>

Qin et al. [129] have calculated the RT TC for single layer phosphorene of 30.15 W/m-K in ZZ direction and 13.65 W/m-K in AC directions in the framework of DFPT and BTE within relaxation time approximation approaches. Authors argued that the minor contribution around 5 % of the ZA mode to total TC is responsible for the low values of TC of phosphorene in comparison with graphene. In contrast, Jain and McGaughey [130] within a similar approach but without RTA for single layer phosphorene have predicted much higher TC values of 110 W/m-K and 36 W/m-K at RT along ZZ and AC directions with an anisotropy ratio of three. The anisotropy was attributed to the anisotropic phonon dispersion, which leads to direction-dependent phonon group velocities. Zhang et al. [131] using NEMD simulations have surprisingly found that TC of few-layer phosphorene (with in-plane dimensions 20 x 20 nm$^2$) is insensitive to the number of layers, which is in strong contrast to the thermal transport behavior in other 2D layered materials. This layer-independent TC was associated with the hindering of flexural phonon mode due to the puckered structure of phosphorene. Authors have also predicted an increase in the TC of phosphorene due to the tensile strain, in particular in the AC direction.

5.  **Conclusions**

We have reviewed phonon and thermal properties of semiconductor nanostructures (multilayer films, core/shell and segmented nanowires), graphene and related two-dimensional materials: hexagonal boron nitride, molybdenum disulfide and black phosphorus. Controlled modification of phonon modes in semiconductor nanostructures opens up new possibilities for phonon-engineered optimization of electrical and heat transport. Unique features of quasi-two-dimensional phonon transport in graphene and related materials are discussed and comparative analysis of experimental and theoretical data on thermal conductivity in single-layer graphene, few-layer graphene, hexagonal boron nitride, molybdenum disulfide and black phosphorus is provided.

**Acknowledgements**

DLN thanks Prof. Alexander A. Balandin (University of California – Riverside) for fruitful discussions of phonon and thermal phenomena in semiconductor nanostructures and graphene. Authors acknowledge the financial support from the Republic of Moldova through the projects 15.817.02.29F and 17.80013.16.02.04/Ua.

**References**

[1] Balandin A A 2005 *J. Nanosci. Nanotechnol.* **5** 1015.

[2] Balandin A A and Nika D L 2012 *Materials Today* **15** 266.

[3] Balandin A A, Pokatilov E P and Nika D L 2007 *J. Nanoel. Optoel.* **2** 140.

[4] Bannov N, Mitin V and Stroscio M 1994 *Phys. Stat. Sol. (b)* **183** 131.

[5] Bannov N, Aristov V, Mitin V and Stroscio M A 1995 *Phys. Rev. B* **51** 9930.

[6] Nishiguchi N, Ando Y and Wybourne M 1997 J. *Phys.: Condens. Matter* **9** 575.

[7] Balandin A A and Wang K L 1998 *Phys. Rev. B* **58** 1544.

[8] Balandin A A and Wang K L 1998 J. *Appl. Phys.* **84** 614.

[9] Khitun A, Balandin A A and Wang K L 1999 *Superlattices Microstruct.* **26** 181.

[10] Zou J and Balandin A A 2001 *J. Appl. Phys.* **89** 2932.

[11] Mingo N 2003 *Phys. Rev. B* **68** 113308.

[12] Mingo N and Yang L 2003 *Phys. Rev. B* **68**, 245406.

[13] Zincenco N D, Nika D L, Pokatilov E P and Balandin A A. 2007 *J. Physics: Conference Series* **92**, 012086.

[14] Aksamija Z and Knezevic I 2010 *Phys. Rev. B* **82** 045319.

[15] Cocemasov A I and Nika D L. 2012 *J. Nanoel. Optoel.* **7** 370.

[16] Crismari D V and Nika D L. 2012 *J. Nanoel. Optoel.* **7** 701.

[17] Cocemasov A I, Nika D L, Fomin V M, Grimm D and Schmidt O G 2015 *Appl. Phys. Lett.* **107** 011904.

[18] Li D, Wu Y, Kim P, Shi L, Yang P and Majumdar A. 2003 *Appl. Phys. Lett.* **83** 2934Ц.

[19] Liu W and Ashegi M 2006 *J. Heat Transfer* **128**, 75 ().

[20] Boukai A I, Bunimovich Y, Tahir-Kheli J, Yu J-K, Goddard III W A and Heath J R 2008 *Nature* **451** 168.

[21] Hochbaum A I, Chen K, Delgado R D, Liang W, Garnett E C, Najarian M, Majumdar A and Yang P 2008 *Nature* **451** 163.

[22] Pokatilov E P, Nika D L and Balandin A A 2003 *Superlattices Microstruct.* **33** 155.

[23] Pokatilov E P, Nika D L and Balandin A A 2004 *Appl. Phys. Lett.* **85** 825.

[24] Pokatilov E P, Nika D L and Balandin A A 2005 *Phys. Rev. B* **72** 113311.

[25] Nika D L, Pokatilov E P, and Balandin A A 2008 *Appl. Phys. Lett.* **93** 173111.

[26] Nika D L, Zincenco N D and Pokatilov E P. 2009 *J. Nanoel. Optoel.* **4** 180.

[27] Nika D L, Pokatilov E P, Balandin A A, Fomin V M, Rastelli A and Schmidt O G 2011 *Phys. Rev. B* **84** 165415.

[28] Hu M, Giapis K P, Goicochea J V, Zhang X and Poulikakos D 2011 *Nano Lett.* **11** 618.

[29] Wingert M C, Chen Z C Y, Dechaumphai E, Moon J, Kim J-H, Xiang J and Chen R 2011 *Nano Lett.* **11** 5507.

[30] Nika D L, Cocemasov A I, Isacova C I, Balandin A A, Fomin V M and Schmidt O G 2012 *Phys. Rev. B* **85** 205439.

[31] Bi K, Wang J, Wang Y, Sha J, Wang Z, Chen M and Chen Y 2012 *Phys. Lett. A* **376** 2668.

[32] Hu M, Poulikakos D 2012 *Nano Lett.* **12** 5487.

[33] Nika D L, Cocemasov A I, Crismari D V and Balandin A A 2013 *Appl. Phys. Lett.* **102** 213109.

[34] Benisty H, Sotomayor-Torrès C M and Weisbuch C 1991 *Phys Rev B* **44**, 10945.

[35] Murdin B N, Heiss W, Langerak C J G M, Lee S-C, Galbraith I, Strasser G, Gornik E, Helm M and Pidgeon C R 1999 *Phys. Rev. B* **55** 5171.

[36] Heitz R, Born H, Guffarth F, Stier O, Schliwa A, Hoffmann A and Bimberg D 2001 *Phys. Rev. B* **64** 241305(R).

[37] Inoue K and Matsuno T 1993 *Phys. Rev. B* **47** 3771.

[38] Tsuchia T and Ando T 1993 *Phys. Rev. B* **48**, 4599.

[39] Zianni X, Simserides C and Triberis G, 1997 *Phys. Rev. B* **55** 16324.

[40] Pokatilov E P, Nika D L, Askerov A S and Balandin A A 2007 *J. App. Phys.* **102** 054304.

[41] Pokatilov E P, Nika D L and Balandin A A 2004 *J. Appl. Phys.* **95** 5625.

[42] Pokatilov E P, Nika D L and Balandin A A 2006 *Appl. Phys. Lett.* **89** 113508.

[43] Fonoberov V A and Balandin A A 2006 *Nano Lett.* **6** 2442.

[44] Kargar F, Debnath B, Kakko J-P, Sainatjoki A, Lipsanen H, Nika D L, Lake R K and Balandin A A 2016 *Nature Comm.* **7** 13400.

[45] Novoselov K S, Geim A K, Morozov S V, Jiang D, Zhang Y, Dubonos S V, Grigorieva I V and Firsov A A 2004 *Science* **306** 666.

[46] Zhang Y, Tan Y-W, Stormer H L and Kim P 2005 *Nature* **438** 201.

[47] Nair R R, Blake P, Grigorenko A N, Novoselov K S, Booth T J, Stauber T, Peres N M R and Geim A K 2008 *Science* **320** 1308.

[48] Mak K F, Shan J and Heinz T F 2011 *Phys. Rev. Lett.* **106** 046401.

[49] Balandin A A, Ghosh S, Bao W, Calizo I, Teweldebrhan D, Miao F and Lau C 2008 *Nano Lett.* **8** 902.

[50] Balandin A A 2011 *Nat. Mat.* **10** 569.

[51] Nika D L and Balandin A A 2017 *Rep. Prog. Phys.* **80** 36502.

[52] Ghosh S, Calizo I, Teweldebrhan D, Pokatilov E P, Nika D L, Balandin A A, Bao W, Miao F. and Lau C N 2008 *Appl. Phys. Lett.* **92** 151911.

[53] Bolotin K I, Sikes K J, Jiang Z, Klima M, Fudenberg G, Hone J, Kim P and Stormer H L 2008 *Solid State Commun.* **146** 351.

[54] Hong X, Posadas A, Zou K, Ahn C H and Zhu J 2009 *Phys. Rev. Lett.* **102** 136808.

[55] Cai W, Moore A L, Zhu Y, Li X, Chen S, Sh L. and Ruoff R S 2010 *Nano Lett.* **10,** 1645.

[56] Jauregui L A, Yue Y, Sidorov A N, Hu J, Yu Q, Lopez G, Jalilian R, Benjamin D K, Delk D A, Wu W, Liu Z, Wang X, Jiang Z, Ruan X, Bao J, Pei S.S and Chen Y.P. 2010 *ECS Trans.* **28** 73.

[57] Chen S, Wu Q, Mishra C, Kang J, Zhang H, Cho K, Cai W, Balandin A A and Ruoff R S 2012 *Nat. Mat.* **11** 203.

[58] Li H, Ying H, Chen X, Nika D L, Cocemasov A I, Cai W, Balandin A A and Chen S 2014 *Nanoscale* **6** 13402.

[59] Dorgan V E, Behnam A, Conley H J, Bolotin K I and Pop E 2013 *Nano Lett.* **13** 4581.

[60] Seol J H, Jo I, Moore A L, Lindsay L, Aitken Z H, Pettes M T, Li X, Yao Z, Huang R, Broido D, Mingo N and Ruoff R S, Shi L 2010 *Science* **328** 213.

[61] Nika D L, Pokatilov E P, Askerov A S and Balandin A A 2009 *Phys. Rev. B* **79** 155413.

[62] Lindsay L, Broido D and Mingo N 2010 *Phys. Rev. B* **82** 115427.

[63] Lindsay L, Li W, Carrete J, Mingo N, Broido D A and Reinecke T L 2014 *Phys. Rev. B* **89** 155426.

[64] Klemens P G 2000 *J. Wide Bandgap Mater.* **7** 332.

[65] Klemens P G 2001 *Int. J. Thermophys.* **22** 265.

[66] Nika D L, Ghosh S, Pokatilov E P and Balandin A A 2009 *Appl. Phys. Lett.* **94** 203103.

[67] Nika D L, Askerov A S and Balandin A A 2012 *Nano Lett.* **12** 3238.

[68] Fugallo G, Cepellotti A, Paulato L, Lazzeri N, Marzari N and Mauri F 2014 *Nano Lett.* **14** 6109.

[69] Munoz E, Lu J and Yakobson B I 2010 *Nano Lett.* **10** 1652-1656

[70] Jiang J-W, Wang J-S and Li B 2009 *Phys. Rev. B* **79** 205418.

[71] Jang Y Y, Cheng Y, Pei Q X, Wang C W and Xiang Y 2012 *Phys. Lett. A* **376** 3668.

[72] Park M, Lee S C and Kim Y S 2013 *J. Appl. Phys.* **114** 053506

[73] Alofi A and Srivastava G P 2012 *J. Appl. Phys.* **112** 013517.

[74] Alofi A and Srivastava G P 2013 *Phys. Rev. B* **87** 115421.

[75] Ma F, Zheng H B, Sun Y J, Yang D, Xu K W and Chu P K 2012 *Appl. Phys. Lett.* **101** 111904.

[76] Fthenakis Z G, Zhu Z and Tomanek D 2014 *Phys. Rev. B* **89** 125421.

[77] Pereira L F C and Donadio D 2013 *Phys. Rev. B* **87** 125424.

[78] Malekpour H, Ramnani P, Srinivasan S, Balasubramanian G, Nika D L, Mulchandani A, Lake R K and Balandin A A 2016 *Nanoscale* **8** 14608.

[79] Shen Y, Xie G, Wei X, Zhang K, Tang M, Zhong J, Zhang G and Zhang Y-W 2014 *J. Appl. Phys.* **115** 063507.

[80] Feng T and Ruan X 2018 *Phys. Rev. B* **97** 045202.

[81] Jang W, Chen Z, Bao W, Lau C N and Dames C 2010 *Nano Lett.* **10** 3909-3913.

[82] Nika D L and Balandin A A 2012 *J. Phys.: Condens. Matter* **24** 233203.

[83] Shahil K M F and Balandin A A 2012 *Nano Lett.* **12** 861.

[84] Pop E, Varshney V and Roy A K 2012 *MRS Bull.* **37** 1273.

[85] Sadeghi M, Pettes M T and Shi L 2012 *Solid State Commun.* **152** 1321.

[86] Wermhoff A P 2012 *Int. J. Trans. Phenom.* **13** 121.

[87] Renteria J D, Nika D L and Balandin A A 2014 *Appl. Sci.* **4** 525.

[88] Malekpour H and Balandin A A 2018 *J. Raman Spectrosc.* **49** 106.

[89] Ghosh S, Bao W, Nika D L, Subrina S, Pokatilov E P, Lau C N and Balandin A A 2010 *Nat. Mater.* **9** 555.

[90] Jo I, Pettes M T, Lindsay L, Ou E, Weathers A, Moore A L and Shi L 2015 *AIP Advances* **5** 053206.

[91] Jeong J Y, Lee K M, Shrestha R, Horne K, Das S, Choi W, Kim M and Choi T Y 2016 *Mater. Res. Express* **3** 055004.

[92] Liu J, Wang T, Xu S, Yuan P, Xu X and Wang X 2016 *Nanoscale* **8** 10298.

[93] Pettes M T, Jo I, Yao Z and Shi L 2011 *Nano Lett.* **11** 1195.

[94] Jang W, Bao W, Ling L, Lau C N and Dames C 2013 *Appl. Phys. Lett.* **103** 133102.

[95] Wang Z, Xie R, Bui C T, Liu D, Ni X, Li B and Thong J T L 2011 *Nano Lett.* **11** 113.

[96] Cocemasov A I, Nika D L, and Balandin A A 2013 *Phys. Rev. B* **88**, 035428.

[97] Nika D L, Cocemasov A and Balandin A A 2014 *Appl. Phys. Lett.* **105** 031904.

[98] Cocemasov A I, Nika D L and Balandin A A 2015 *Nanoscale* **7** 12851.

[99] Limbu T B, Hahn K R, Mendoza F, Sahoo S, Razink J J, Katiyar R S and Morell G 2017 *Carbon* **117** 367.

[100] Kuang Y, Lindsay L and Huang B 2015 *Nano Lett.* **15** 6121.

[101] Zhan H, Zhang Y, Bell J M and Gu Y 2015 *J. Phys. Chem. C* **119** 1748.

[102] Shen X, Wang Z, Wu Y, Liu X, He Y B and Kim J K 2016 *Nano Lett.* **16** 3585.

[103] Si C, Lu G, Cao B Y, Wang X D, Fan Z and Feng Z H 2017 *J. Appl. Phys.* **121** 205102.

[104] D'Souza R and Mukherjee S 2017 *Phys. Rev. B* **95** 085435.

[105] Li L H and Chen Y 2016 *Adv. Funct. Mater.* **26** 2594.

[106] Jo I, Pettes M T, Kim J, Watanabe K, Taniguchi T, Yao Z and Shi L 2013 *Nano Lett.* **13** 550.

[107] Zhou H, Zhu J, Liu Z, Yan Z, Fan X, Lin J and Tour J M 2014 *Nano Res.* **7** 1232.

[108] Wang C, Guo J, Dong L, Aiyiti A, Xu X and Li B 2016 *Sci. Rep.* **6** 25334.

[109] Sichel E K, Miller R E, Abrahams M S and Buiocchi C J 1976 *Phys. Rev. B* **13** 4607.

[110] Lindsay L and Broido D A 2011 *Phys. Rev. B* **84** 155421.

[111] Mak K F, Lee C, Hone J., Shan J and Heinz T F. 2010 *Phys. Rev. Lett.* **105** 136805.

[112] Radisavljevic B, Radenovic A, Brivio J, Giacometti I V and Kis A 2011 *Nat. Nanotechnol.* **6** 147.

[113] Yan R, Simpson J R, Bertolazzi S, Brivio J, Watson M, Wu X and Xing H G 2014 *ACS Nano* **8** 986.

[114] Jo I, Pettes M T, Ou E, Wu W and Shi L 2014 *Appl. Phys. Lett.* **104** 201902.

[115] Taube A, Judek J, Łapińska A and Zdrojek M 2015 *ACS Appl. Mater. Interfaces* **7** 5061.

[116] Cai Y, Lan J, Zhang G and Zhang Y W 2014 *Phys. Rev. B* **89** 035438.

[117] Ding Z, Jiang J W, Pei Q X and Zhang Y 2015 *Nanotechnology* **26** 065703.

[118] Ding Z, Pei Q X, Jiang J W and Zhang Y W 2015 *J. Phys. Chem. C* **119** 16358.

[119] Gu X, Li B and Yang R 2016 *J. Appl. Phys.* **119** 085106.

[120] Liu J, Choi G M and Cahill D G 2014 *J. Appl. Phys.* **116** 233107.

[121] Li L, Yu Y, Ye G J, Ge Q, Ou X, Wu H and Zhang Y 2014 *Nat. Nanotechnol.* **9** 372.

[122] Xia F, Wang H and Jia Y 2014 *Nat. Commun.* **5** 4458.

[123] Koenig S P, Doganov R A, Schmidt H, Castro Neto A H and Özyilmaz B 2014 *Appl. Phys. Lett.* **104** 103106.

[124] Luo Z, Maassen J, Deng Y, Du Y, Garrelts R P, Lundstrom M S and Xu X 2015 *Nat. Commun.* **6** 8572.

[125] Jang H, Wood J D, Ryder C R, Hersam M C and Cahill D G 2015 *Adv. Mater.* **27** 8017.

[126] Lee S, Yang F, Suh J, Yang S, Lee Y, Li G and Park J 2015 *Nat. Commun.* **6** 8573.

[127] Hong Y, Zhang J, Huang X and Zeng X C 2015 *Nanoscale* **7** 18716.

[128] Liu T H and Chang C C 2015 *Nanoscale* **7** 10648.

[129] Qin G, Yan Q B, Qin Z, Yue S Y, Hu M and Su G 2015 *Phys. Chem. Chem. Phys.* **17** 4854.

[130] Jain A and McGaughey A J 2015 *Sci. Rep.* **5** 8501.

[131] Zhang Y Y, Pei Q X, Jiang J W, Wei N and Zhang Y W 2016 *Nanoscale* **8** 483.

**FIGURE CAPTIONS**

**Figure 1.** Energy dispersions of the dilatational (SA) acoustic phonons in 2-nm thick free-standing Si slab (a), Diamond/Si/Diamond heterostructures with dimensions 1 nm/ 2 nm/ 1 nm (b) and 4 nm/ 2 nm/ 4 nm (c) and 2-nm thick Si slab with clamped boundaries (d). Reprinted with permission from Ref. [25] (copyright 2008 American Institute of Physics).

**Figure 2**. Average phonon group velocity as a function of the phonon frequency for dilatational modes in rectangular GaN, GaN/AlN and GaN/Plastic nanowires. Reprinted with permission from Ref. [24] (copyright 2005 American Physical Society).

**Figure 3**. Temperature dependence of phonon thermal conductivity in Si nanowires and Si/SiO2, Si/Pl and Si/Diamond core/shell nanowires.

**Figure 4**. Phonon thermal conductivity as a function of the absolute temperature. Results are presented for Si NW, Si SMNW and core/shell Si/Ge SMNWs with different thickness of Ge. Reprinted with permission from Ref. [33] (copyright 2013 American Institute of Physics).

**Figure 5.** Measured and calculated phonon dispersions for a GaAs nanowire with diameter of 122 nm along [111] direction. Reprinted with permission from Ref. [44] (copyright 2016 Nature Publishing Group).

**Figure 6.** (a) Thermal conductivity of graphene as a function of temperature shown for different graphene flake widths. The results were calculated for the specularity parameter $p$=0.9. An experimental data point after Refs. [49, 52] is also shown for comparison. (b) Thermal conductivity of graphene as a function of the density of defects induced by electron beam irradiation. The experimental results are shown by square, circle and triangle points. The solid curves correspond to the BTE results, plotted for different values of specularity parameter $p$. Adopted with permission from Ref. [61] (copyright 2009 American Physical Society) and from Ref. [78] (copyright 2016 The Royal Society of Chemistry).

**Figure 7**. (**a**) Thermal conductivity of suspended SLG, AB-BLG and T-BLG as a function of the measured temperature; (**b**) Dependence of the difference of the specific heat in T-BLG from that in AB-BLG on the temperature. The inset shows the relative difference between AB-BLG and T-BLG specific heats as a function of temperature. Reprinted with permission from Ref. [58] (copyright 2014 The Royal Society of Chemistry) and from Ref. [97] (copyright 2014 American Institute of Physics).

**Figure 8**. Orientation-dependent TC of 1 μm-length phosphorene sheet. The gray dot line is fitted by cosine function with period 2π. The ZZ and AC chiralities correspond to 0 and 90 degrees, respectively. Reproduced with permission from Ref. [128] (copyright 2015 The Royal Society of Chemistry).

**FIGURES**

![Figure 1: Phonon dispersion plots for (a) 2 nm free-standing Si slab, (b) Diamond/Si/Diamond 1 nm/2 nm/1 nm, (c) Diamond/Si/Diamond 4 nm/2 nm/4 nm, and (d) 2 nm clamped Si slab.]

**Figure 1 of 8. A. I. Cocemasov et al.**

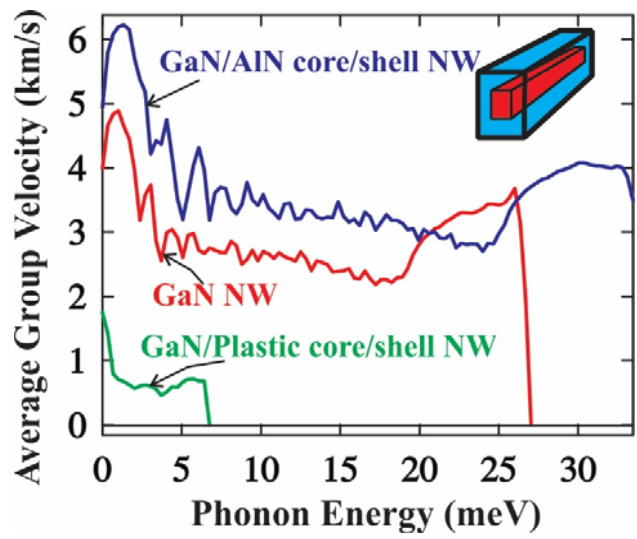

**Figure 2 of 8. A. I. Cocemasov et al.**

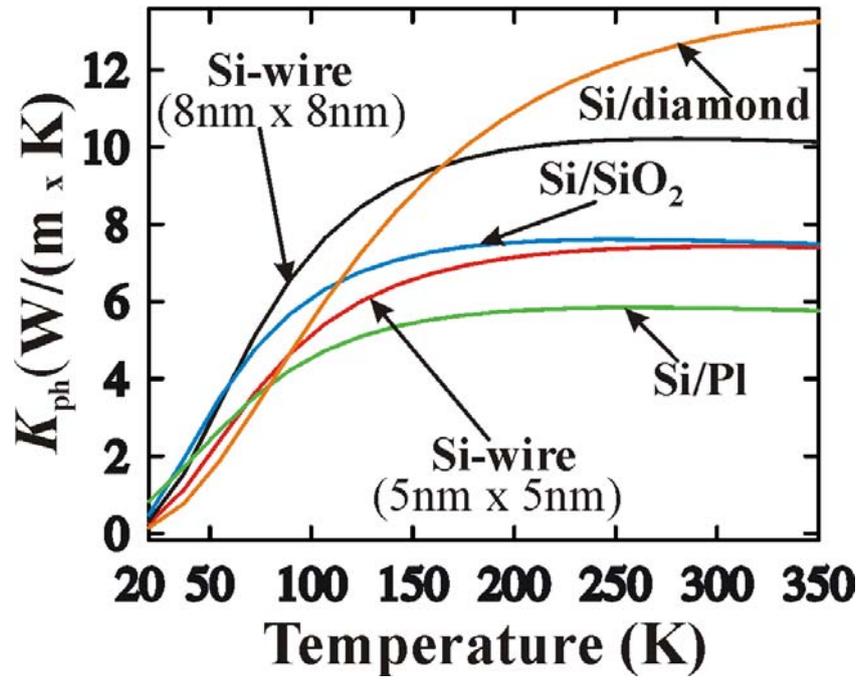

**Figure 3 of 8. A. I. Cocemasov et al.**

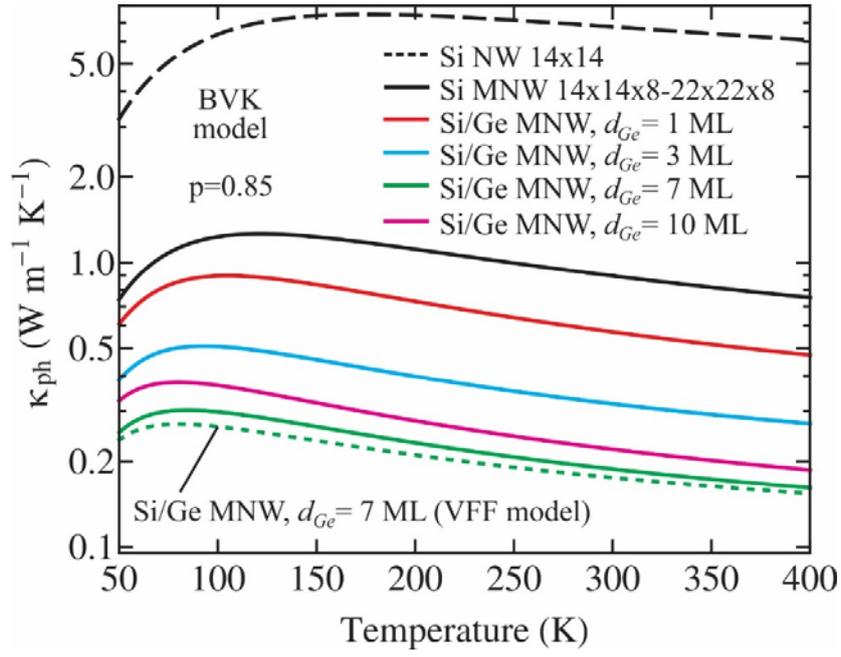

**Figure 4 of 8**. **A. I. Cocemasov et al**.

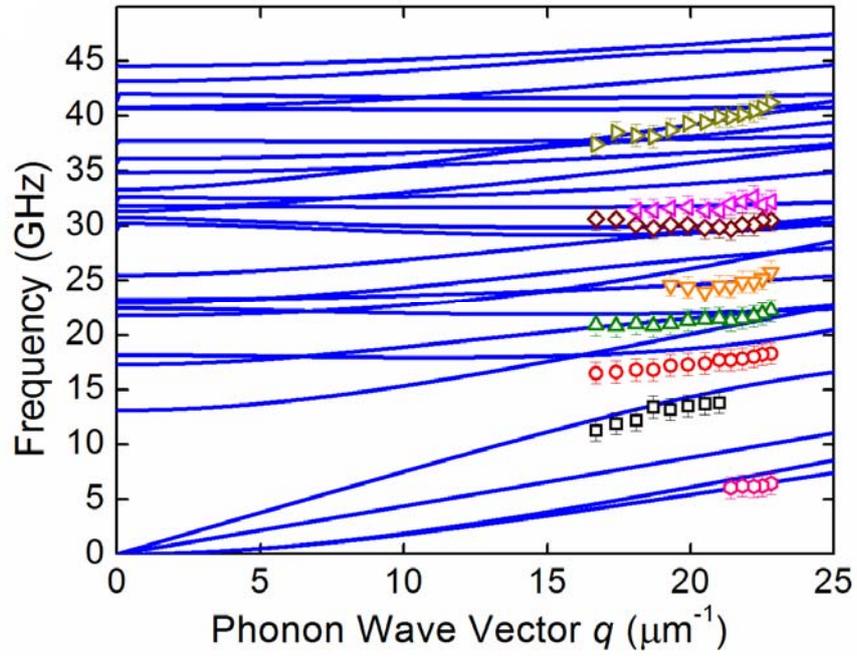

**Figure 5 of 8. A. I. Cocemasov et al**.

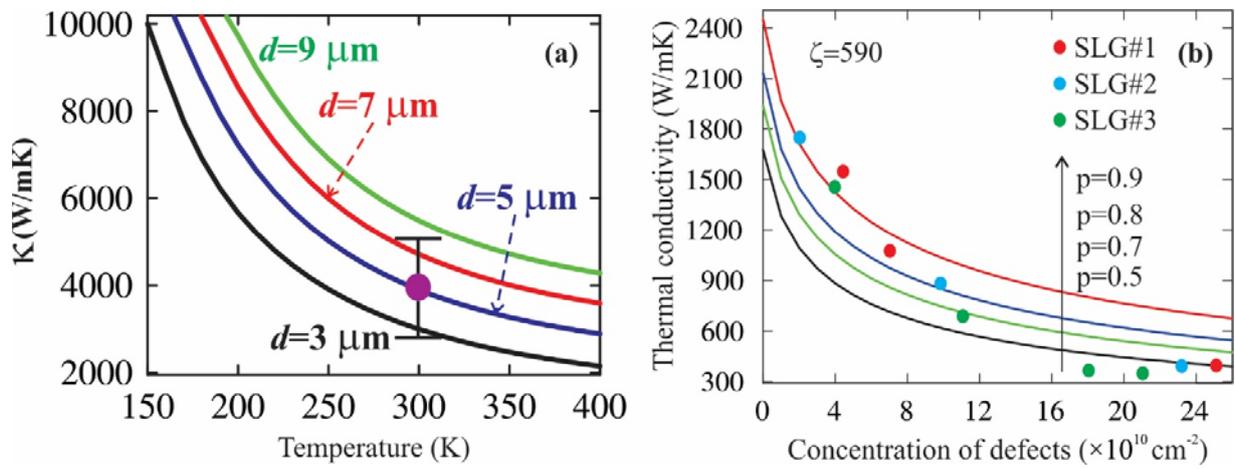

**Figure 6 of 8. A. I. Cocemasov et al**.

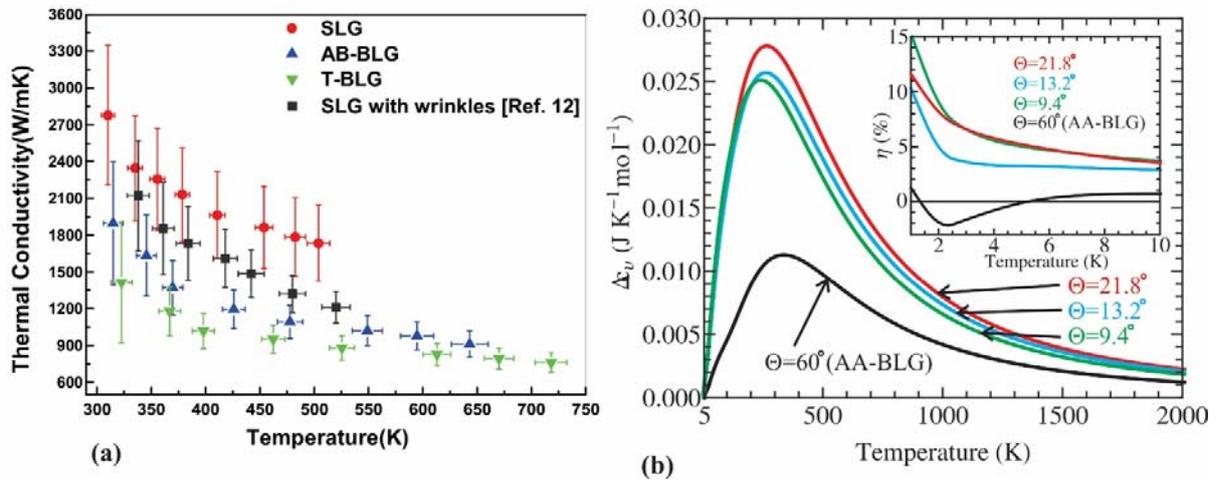

**Figure 7 of 8. A. I. Cocemasov et al**.

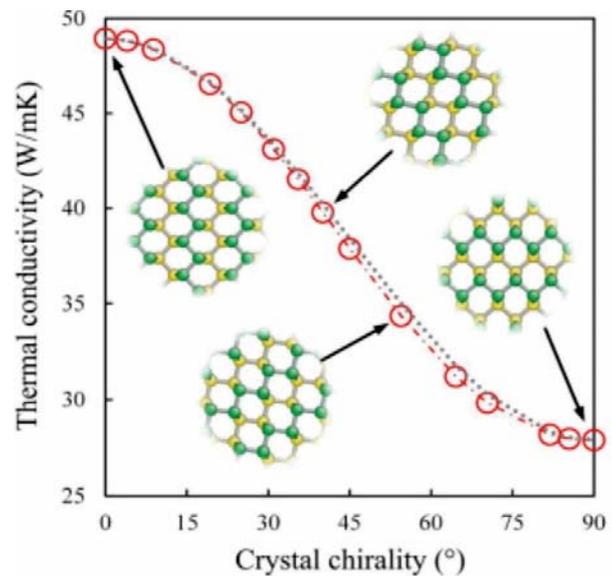

**Figure 8 of 8. A. I. Cocemasov et al**.

**Table 1.** Thermal conductivity of single-layer graphene: experimental data

| κ (W/mK) | Method | Brief description | Ref. |
|---|---|---|---|
| ~3000 – 5000 | Raman optothermal | suspended; exfoliated; $T$~300 K | 49,52 |
| 2500 | | suspended; chemical vapor deposition (CVD) grown; $T$~300 K | 55 |
| 1500 – 5000 | | suspended; CVD grown | 56 |
| 1600 – 2800 | | suspended; strong isotope dependence; $T$~380 K | 57 |
| 2778.3 ± 569 | | suspended, $T$ ~ 325 K | 58 |
| 2000 – 3800 | electrical self-heating | exfoliated and CVD grown; $T$~300 K | 59 |
| 600 | electrical | supported; exfoliated | 60 |

**Table 2.** Thermal conductivity of single-layer graphene: theoretical data

| Brief description | Refs. |
|---|---|
| Strong dependence of TC on flake size | 61 - 72 |
| Strong dependence of TC on strain | 73 - 75 |
| Strong dependence of TC on point defects | 71, 73, 74, 76, 77 |
| Strong dependence of TC on isotopes | 73, 74, 78 |
| Strong dependence of TC on edge roughness | 61, 67, 77, 79 |